\documentstyle[12pt,epsfig]{article}

\def\ee{\end{equation}}
\def\be{\begin{equation}}
\def\eea{\end{eqnarray}}
\def\bea{\begin{eqnarray}}
\def\eeas{\end{eqnarray*}}
\def\beas{\begin{eqnarray*}}
 
\begin{document}
\pagenumbering{arabic}
\title{The solar neutrino spectrum derived from electron scattering and 
charged current interactions}
\author{G. Fiorentini and F.L. Villante}

\date{}
 
\maketitle
 
\small{Dipartimento di Fisica, Universit\`a di Ferrara, I-44100,
Ferrara (Italy)}
 
\small{Istituto Nazionale di Fisica Nucleare, Sezione di Ferrara,
I-44100, Ferrara (Italy)}
 
\bigskip
 
\begin {abstract}  
We provide a method for  extracting information on the energy spectrum
of solar neutrinos directly  from the spectrum of scattered electrons.
As an example, we apply it to the published Super-Kamiokande data. 
When combined with data from SNO on charged current interactions this 
method allows to derive separately the spectra of $\nu_{e}$ and of 
$\nu_{\mu}$ plus $\nu_{\tau}$. 
\end{abstract}
\smallskip
%\noindent PACS number(s): ......
 
\bigskip
\bigskip
\bigskip
\bigskip
\noindent\small{\underline{Corresponding author}\\
F.L. Villante\\
Dipartimento di Fisica\\
Via Paradiso, 12 - 44100 Ferrara (Italy)\\
Telephone Number: +39 0532781879\\
Fax Number: +39 0532762057\\

\pagebreak

\section{Introduction}

As a general rule, important information is contained in the
spectrum of the radiation which is being used as a probe of a physical
system. This holds for any radiation and thus also for neutrinos
emitted from the Sun. Indeed, the reconstruction of the solar neutrino
spectrum from experimental data is very interesting, as a way to
establish neutrino properties and/or to study the stellar interior.

Solar neutrino experiments provide however indirect information about
the neutrino spectrum. As an example, from  radiochemical experiments
\cite{Hom,Gal,Sage}, which
detect the decay of nuclei produced  by neutrino capture over the
target atoms, one can derive an average flux of electron neutrinos
arriving onto earth, where the average is weighted with the  capture
cross section and it generally involves neutrinos from different branches
of the fusion chain. 

Super-Kamiokande (SK) \cite{sk} is essentially sensitive to  $^{8}$B neutrinos only. 
It studies Electron Scattering (ES), 
%.............................................................................
\begin{equation}
\nu + e^- \to \nu + e^-\ ~,
\label{nu_e}
\end{equation}
%.............................................................................
by measuring the energy distribution of electrons for total 
energies $E_e > 5.0$ MeV. Electron scattering is being studied 
also by SNO, see \cite{sno}, for kinetic energies $T_{e} > 6.75$ MeV. 

ES measurements have  given, so far, only indirect information on the
distribution of the neutrino energy $E_{\nu}$.  In the usual approach, one
starts with a flux $\varphi_{a}(E_{\nu})$, where the index $a$ 
corresponds to active neutrinos, and evaluates the scattered 
electron spectrum. If this is
consistent (inconsistent) with experimental data
then the input $\varphi_a$ is
accepted (excluded) for describing the solar neutrino flux arriving onto
Earth.      

The main purpose of this letter is to provide an alternative method for
extracting the energy spectrum of solar neutrinos directly from data on
the spectrum of scattered electrons.

The basic idea is very simple. Electrons with total energy $E_e$  (which
we assume much larger than $m_e$) are produced by neutrinos with 
$E_{\nu} > E_e - m_e/2$.
Above this threshold,  the cross sections $d \sigma_a / dE_e$ are
practically independent of electron energy. Thus the
difference of  electron spectra centered  
at two neighbouring  energies around $E_e$  
gets contribution only from neutrinos
 with energy around $E_{\nu}= E_e - m_e/2$.  Consequently, the
neutrino spectrum can be calculated from the derivative of the measured
electron energy spectrum. 

First, we shall present this scheme by using simple analytical
approximations. Next we shall confirm these estimates by means of a full
numerical calculation.

As an example, we shall apply this method to the published SK data~\cite{sk},
 see our results in fig.~\ref{fig5}.
 We recommend however that a full analysis be done by the
experimental group, since a detailed knowledge of the detector and of 
the data is important for extracting optimal information.

From ES experiments one derives information on a specific combination
of the spectrum of $\nu_{e}$, $\varphi_{e}$, with the spectrum of $\nu_{\mu}$
plus $\nu_{\tau}$, $\varphi_{\mu\tau}$:
\be
\varphi_{\rm ES}=\varphi_{e} + \beta \varphi_{\mu\tau}
\ee
where $\beta$ is a suitable ratio of $\nu_{\mu}$ ($\nu_{\tau}$) 
to $\nu_{e}$ cross sections.
  An important addition of  SNO \cite{sno}, by means of the Charged Current (CC) data on
deuterium, 
\be
\nu_{e} + d \rightarrow p + p + e^{-} ~,
\label{cc_reac}
\ee
is the possibility of determining  the $\nu_{e}$ spectrum.
 We will show that, by combining ES and CC data, 
one can determine  separately the spectrum of 
active neutrinos different from $\nu_e$.

We remind that  the comparison of the (energy integrated) 
electron signal  from ES  and CC data has already allowed to disentangle 
the presence of a ${\nu}_{\mu\tau}$ component in the  electron scattering data
\cite{sno,sksno,vari}.
 By the method which we are proposing it becomes possible to determine
the $\nu_{\mu\tau}$ energy spectrum separately. This is not only interesting
by itself, but it also offers a possibility to find a model-independent 
signature of sterile neutrinos in case they too are 
coming from the sun, as we shall comment at the end of the paper.

\section{From electron spectrum to the neutrino spectrum}

{\em a) A simplified case}\\
In the limit of infinite energy resolution and assuming full efficiency
for electron detection, the electron energy spectrum from reaction
(\ref{nu_e}) is given by:
\be
\frac{dN}{dE_e} = {\mathcal N} \, t \, \int dE_{\nu} \;
\left[ \varphi_e(E_{\nu}) \frac{d\sigma_e}{dE_{e}} (E_{\nu},E_{e}) 
+ \varphi_{\mu\tau}(E_{\nu}) \frac{d\sigma_{\mu}}{dE_{e}} (E_{\nu},E_e)
\right]~,
\label{sk_spec_nr}
\ee
where ${\mathcal N}$ is the number of target electrons, $t$ is the 
measurement time, 
$\varphi_a$ are the neutrino fluxes and  
$d\sigma_a/dE_{e}$ are the differential cross sections for active 
neutrinos ($a =e,\mu,\tau$) \cite{cross_sk,cross_sk2}.

For energies much larger than $m_e$, electrons with total energy
$E_{e}$ are produced by neutrinos with energy larger than  
$E_{\min} = E_e - m_e/2$.
In addition, to a good approximation in the range of measured electron energies,
the differential cross sections
$d\sigma_a/dE_{e}$ are independent of the energy of
the scattered electron. In other words, we can write:
\be
\frac{d\sigma_{a}}{dE_{e}} 
= \sigma_{a}(E_{\nu}) \, \theta(E_{\nu}-E_{\rm min}) ~,
\label{cross_app}
\ee
where the factors $\sigma_{a}$ are, moreover, weakly dependent on the
neutrino energy $E_{\nu}$.
Therefore, by differentiating both sides of eq.
(\ref{sk_spec_nr}) with respect to the electron energy 
we obtain:
\be
\varphi_e(E_{\nu}) + \beta \varphi_{\mu\tau}(E_{\nu}) 
=  - \left( \frac{1}{{\mathcal N} \,t \, \sigma_e}\; \right) 
\frac{d^{2} N}{dE_e ^2} ~.
\label{spec_der_nr}
\ee
The factor $\beta = \sigma_{\mu}/\sigma_{e}$  is  weakly dependent on 
$E_\nu$ in the energy of interest to us and the r.h.s. is calculated at:
\be
E_{e} = E_{\nu} + m_e / 2 ~.
\label{ee_enu_nr}
\ee

Eq. (\ref{spec_der_nr}), which is our basic result, is the direct  information on the
neutrino energy spectrum which  can be derived from the electron
spectrum. We remark that the r.h.s is purely defined in terms of measurable
quantities.

 Of course, there is no way at this stage to tell which are
the separate contributions of $\nu_e$ and $\nu_{\mu\tau}$. 
This separation requires an additional information, such as it 
can be provided by the CC measurement of SNO, see below. 

On the other hand equation (\ref{spec_der_nr}) allows a direct test of oscillation
models. Any proposed oscillation solution 
predicts a definite expression for the 
produced flux $\varphi(E_{\nu})$ and for the 
oscillation probabilities $P_{ea}(E_{\nu})$. 
In terms of these, the l.h.s of eq.~(\ref{spec_der_nr}) can be
immediately calculated  as $\varphi\,[P_{ee}+\beta (P_{e\mu}+P_{e\tau})]$ 
and compared with the observable quantity on the r.h.s. .\\

{\em b) The effect of the finite energy resolution}\\
In practice, the situation is more complicated, since the experiment has
a finite energy resolution. This implies that the observed electron
energy $\epsilon_e$ is different from the true electron energy $E_{e}$. 
The {\em observed} spectrum $S(\epsilon_e)$ is related to the {\em true} energy 
spectrum $dN/dE_e$  by means of the
following relation:
\begin{equation}
S(\epsilon_e) = \int dE_e
\;r(\epsilon_e,E_e) \, \frac{dN}{dE_e} 
\label{true_spec}
\end{equation}
where the resolution function can be taken as a gaussian:
\be            
r(\epsilon_e, E_e) = \frac{1}{\sqrt{2\pi}\Delta} 
\exp\left(-\frac{(\epsilon_{e}-E_{e})^2}
{2\Delta^2}\right)~.
\label{resolution}
\ee
By deriving both sides of eq. (\ref{true_spec}) with respect to the observed 
energy $\epsilon_{e}$ 
we obtain: 
\be
\frac{dS}{d\epsilon_{e}} = \int dE_{e} \; 
\frac{\partial r(\epsilon_{e},E_{e})}{\partial \epsilon_{e}} \; \frac{dN}{dE_{e}}~.
\label{true_der1}
\ee
If we neglect the energy dependence of $\Delta$ 
, from (\ref{resolution})
we have $\partial r/\partial \epsilon_{e} = - \partial r/ \partial E_{e} $. 
Integration by parts gives:
\begin{equation}
\frac{dS}{d\epsilon_{e}} = \int dE_{e} \; 
r(\epsilon_{e},E_{e}) \; \frac{d^{2}N(E_{e})}{dE_{e}^2}~.
\label{true_der2}
\end{equation}
By using eq.(\ref{spec_der_nr}) we obtain:
% ------------------------------------
\be
\frac{dS}{d\epsilon_e} = 
- {\mathcal N} \, t \, \sigma_{e} \int dE_{\nu} 
\;r(\epsilon_e,E_{\nu}+ m_{e}/2)
\left[\varphi_e(E_{\nu}) + \beta \varphi_{\mu\tau}(E_{\nu}) \right]~,
\label{true_der}
\ee
 where we have taken advantage of the weak energy dependence of 
$\sigma_{e}(E_{\nu})$ to take it out from the sign of integration.

Since experimental results are generally presented in terms of the
ratios to the SSM prediction, it is convenient to write eq. (\ref{true_der}) 
in a slightly different form.

We note that the electron spectrum predicted by Standard Solar Model
(SSM) calculations, $dS^{(SSM)}/d\epsilon_e$, satisfies an equation 
similar to (\ref{true_der}):
\be
\frac{dS^{(SSM)}}{d\epsilon_e} = - {\mathcal N} \, t \, \sigma_{e}  
\int  dE_{\nu} 
\;r(\epsilon_e,E_{\nu}+m_{e}/2)
\varphi^{SSM}(E_{\nu}) ~. 
\label{true_der_ssm}
\ee
Also, for each energy we can normalize the neutrino fluxes to the
SSM prediction, $\varphi^{SSM}$, by introducing the quantities:
\be
f_a(E_{\nu}) =   
\frac{\varphi_a(E_{\nu})}{\varphi^{SSM}(E_{\nu})}  ~.
\label{fa}
\ee
With these definitions, from eqs (\ref{true_der}) and (\ref{true_der_ssm}) we have:
\be
\int dE_{\nu}  
\;\left[ f_e(E_{\nu}) + \beta f_{\mu\tau}(E_{\nu}) \right]\;
\rho(\epsilon_e,E_{\nu})=                      
\frac{dS/d\epsilon_e}{dS^{(SSM)}/d\epsilon_e}
\label{der_ratio}
\ee
where the response function $\rho(\epsilon_e,E_{\nu})$ 
is given by:
\be
\rho(\epsilon_{e},E_{\nu}) = \frac{ r(\epsilon_{e},E_{\nu}+m_e /2) \; \varphi^{SSM}(E_{\nu})}
{\displaystyle
\int dE_{\nu}\; r(\epsilon_{e},E_{\nu}+m_e/2) \; \varphi^{SSM}(E_{\nu})} ~.
\label{rho}  
\ee
 The function $\rho(\epsilon_e,E_{\nu})$ essentially 
measures the energy resolution which can be attained for 
determining the neutrino spectrum (or, more precisely, its deviation
from the SSM prediction).

Eq.(\ref{der_ratio}), which is our main result, is the extension of (\ref{spec_der_nr}) 
for the case of finite energy resolution. It has a natural interpretation:
due to the finite energy resolution, the derivative of the observed  electron spectrum
determines the neutrino spectrum, smeared over the energy resolution.

 The relation between neutrino and electron energy,
 previously given by eq.(\ref{ee_enu_nr}), can now be derived as follows. 
For a fixed electron energy $\epsilon_{e}$, the l.h.s of eq.(\ref{der_ratio})
receives contribution from neutrino energies around $E_{\nu}$
such that $ \partial \rho / \partial E_{\nu} = 0$. 
Neglecting again the energy dependence of $\Delta$ this gives:
\be
\epsilon_{e} = E_{\nu} + m_{e}/2 - 
\Delta^2 \;\frac{ \partial \ln \varphi^{SSM}} {\partial E_{\nu}} ~.
\label{ee_enu_app}  
\ee
\\

{\em c) The general case}\\
We remind that equations (\ref{der_ratio}), (\ref{rho}) and (\ref{ee_enu_app})
have been obtained by neglecting the
electron energy dependence of $d \sigma_{a}/d E_{e}$ and of 
$\Delta$. If these approximations are released one obtains: 
\be
\int dE_{\nu}\;
\left[ f_e \, \rho_{e} + \beta f_{\mu\tau} \, \rho_{\mu} \right] =
\frac{dS/d\epsilon_e}{dS^{(SSM)}/d\epsilon_e}
\label{der_ratio_gen}
\ee
where one has now two response functions:
\be
\rho_{a}(\epsilon_{e},E_{\nu}) =
\frac{\displaystyle 
\varphi^{SSM}(E_{\nu}) \int dE_{e}
\frac{\partial r(\epsilon_{e},E_{e})}{\partial \epsilon_{e}} \; 
\frac{d\sigma_{a}(E_{\nu},E_{e})}{dE_{e}}
}
{\Psi_{a}
}~.
\label{rho_a}
\ee
The normalizations $\Psi_{a}$ are given by:
\be
\Psi_{a} =
\int dE_{\nu}\;
\varphi^{SSM}(E_{\nu}) \int dE_{e} \;
\frac{\partial r(\epsilon_{e},E_{e})}{\partial \epsilon_{e}} \; 
\frac{d\sigma_{a}(E_{\nu},E_{e})}{dE_{e}}~.
\label{psi_a}
\ee
The factor $\beta$ is expressed as:
\be
\beta(\epsilon_{e})=
\frac{\Psi_{\mu}}
{\Psi_{e}}~.
\label{beta}
\ee

We can show that for SK eqs.~(\ref{der_ratio}-\ref{ee_enu_app})
are quite accurate. In fact,
in the case of SK, the energy resolution can be described by the
expression~(\ref{resolution}) with $\Delta$ given by \cite{sk_res}:
\be
\Delta = 1.5\; {\rm MeV} \sqrt{(E_{e}-m_{e})/10 {\rm MeV}} 
\label{delta_sk}
\ee
By using this relation, the exact expression 
for the differential cross sections $d\sigma_{a}/dE_{e}$ \cite{cross_sk2}
and the SSM neutrino flux $\varphi^{SSM}$\footnote{We remark that
the response functions depend only on the shape of the neutrino 
spectrum. SK is essentially sensitive to $^8$B neutrinos only,
with a small contribution
from {\it hep} neutrinos. We use the $^{8}$B neutrino spectrum given in 
\cite{8B_spectrum}, and the ratio between $^{8}$B and {\it hep} neutrino flux
predicted by \cite{ssm}. Variations of the {\it hep} 
neutrino flux within the current phenomenological limits do not 
affect our results.},
one can calculate numerically the response functions 
$\rho_{e}$ and $\rho_{\mu}$ for Super-Kamiokande. 
In fig.~\ref{fig1} we present our results as a function of neutrino energy, 
for selected values of $\epsilon_e$. The functions $\rho_{e}$ and $\rho_{\mu}$
are bell shaped functions, with a full-width-half maximum of about 2.5 MeV. 
They are slightly narrower than the electron energy 
resolution $r$, due to the presence of $\varphi^{SSM}$ in their definition. 

One sees that, a part for the smallest electron energies, the exact
response functions $\rho_{e}$ and $\rho_{\mu}$ are close to the approximate
expression $\rho$ given by eq.(\ref{rho}). For our purposes we can thus
safely assume:
\be 
\rho_{e}(\epsilon_{e},E_{\nu})
= \rho_{\mu}(\epsilon_{e},E_{\nu}) = \rho(\epsilon_{e},E_{\nu})~.
\label{rho_equality}
\ee

In fig.~\ref{fig2} we show the relationship between neutrino and electron energies,
calculated numerically by solving 
$\partial \rho_{a}/\partial E_{\nu} = 0$. One sees that the approximate
solution given by eq. (\ref{ee_enu_app}) is quite accurate.

Finally, in fig.~\ref{fig3} we present an estimate for
$\beta$, calculated according to eq.(\ref{beta}),
as a function of the electron energy. As the energy increases, it becomes
approximatively constant, $\beta\simeq 0.15$.

%%%%%%%%%%%%%%%%%%%%%%%%%%%%%%%%%%%%%%%%%%%%%%%%%%%%%%%%%%%%%%%%%%%%%
\section{Extraction  of the $\nu$ energy spectrum from SK data}
%%%%%%%%%%%%%%%%%%%%%%%%%%%%%%%%%%%%%%%%%%%%%%%%%%%%%%%%%%%%%%%%%%%%%

As an example, we outline here a numerical procedure for
extracting physical information on $f_{a}(E_{\nu}) = \varphi_{a}/\varphi^{SSM}$ 
from the published SK data \cite{sk}
by means of eqs. (\ref{der_ratio}-\ref{ee_enu_app}).

We remind that SK measures 
the electron spectrum $S(\epsilon_{e})$ for electron energies in the
range 5-20 MeV, see fig.~\ref{fig4}. Data are grouped into 18 bins, each 0.5 MeV wide, 
covering the range 5-14 MeV, with an additional bin extending 
from 14 to 20 MeV. The signal in the $i$-th bin, $S_{i}$, is associated with
an error $\Delta S_{i}$, which we take (for the moment)
as the sum in quadrature of the statistical and uncorrelated systematical
errors given in \cite{sk}. Data correspond to 1258 days of exposure \cite{sk}.
 All quantities shown in fig.~\ref{fig4} are normalized to the
predictions $S_{i}^{(SSM)}$ corresponding to the Standard Solar Model (SSM)
of \cite{ssm}\footnote{
The SSM of \cite{ssm} predicts a $^{8}$B neutrino fluxes $\Phi_{\rm B} = 
5.05 \cdot 10^6 {\rm cm}^{-2}\,{\rm s}^{-1}$. Here and in the following, 
we use this value as a (convenient) normalization factor. 
We remark however that our results do not depend at all on solar
models.}

In order to evaluate the r.h.s. of eq.(\ref{der_ratio}) and the 
associated error, it is convenient to divide the energy range
probed by SK in a few intervals, each with a width 
comparable to the neutrino energy resolution, since the experiment 
is anyhow unsensitive to spectral 
deformations on smaller scales. On the other hand, by considering intervals
larger than the original SK binning it is possible to decrease statistical
fluctuations in the evaluation of the spectrum derivative.

We have considered five intervals, with energies in MeV between (5-7),
 (7-9), (9-11), (11-13) and (13-20) respectively.   
Inside each interval $S(\epsilon_e)$ and  $S^{(SSM)}(\epsilon_e)$ 
can be approximated by straight lines and their slopes are 
determined by least squares fitting:
\be
\frac{dS}{d\epsilon_e} = 
\frac{\langle S \, \epsilon_e \rangle - 
\langle S \rangle \langle \epsilon_e \rangle}
{\langle \epsilon_e^2 \rangle - \langle \epsilon_e \rangle^2}
\label{der_spe}
\ee
and similarly for $dS^{(SSM)}/d\epsilon_{e}$.
In the average $\langle\;\rangle$ each point is weighted 
according to the experimental error $\Delta S_i$:
\be
\langle X \rangle = \frac{\sum_{i} X_i / \Delta{S_i}^2 } 
{\sum_{i}  1/\Delta S_i^2} ~.
\label{averages}
\ee
For each interval we get
\be
\frac{dS/d\epsilon_e}{dS^{(SSM)}/d\epsilon_e} =
\frac
{\langle S \, \epsilon_e \rangle - 
\langle S \rangle \langle \epsilon_e \rangle}    
{\langle S^{(SSM)} \epsilon_e \rangle - 
\langle S^{(SSM)} \rangle \langle \epsilon_e \rangle}
\label{der_ratio_spe}
\ee
with an associated error:
\be
\delta  = 
\frac
{\left[ \langle \epsilon_e^{2} \rangle - 
\langle \epsilon_e \rangle^{2}\right]^{1/2}}    
{\langle S^{(SSM)} \epsilon_e \rangle - 
\langle S^{(SSM)} \rangle \langle \epsilon_e \rangle}
\;\cdot
\sqrt{\frac{1}
{\sum_{i}  1/\Delta S_i^2}}
~.
\label{err_spe}
\ee

At this point we can calculate, by means of 
eqs.~(\ref{der_ratio}-\ref{ee_enu_app})
the deviations of the neutrino spectrum from the SSM
predictions by using the SK data.

Our results are presented in fig.~\ref{fig5} as a function of the 
neutrino energy. The quantity on the vertical axis represents
$f_{e}+\beta f_{\mu\tau}$ averaged with the
response function $\rho$, i.e.:
\be
F_{\rm ES}(E_{\nu}) = 
\int dE'_{\nu}  
\;\left[ f_e(E'_{\nu}) + \beta f_{\mu\tau}(E'_{\nu}) \right]\;
\rho(\epsilon_e,E'_{\nu})
\label{Fsk}
\ee
where $\epsilon_{e}(E_{\nu})$ is given by eq.(\ref{ee_enu_app}) and 
(\ref{delta_sk}). 
The horizontal bar 
corresponds to the neutrino energy resolution calculated as
the full width half maximum of the response function 
$\rho( \epsilon_{e}, E_{\nu})$.

Concerning errors, the inner vertical bar takes into account statistical
and energy uncorrelated systematical errors. The outer bar also accounts
for energy correlated systematical errors, as given in \cite{sk}.
Their effect  has been evaluated by a simultaneous up and down 
shift of all the experimental points.

We remark that from SK we have been able to derive only a specific 
combination of $\nu_e$ and $\nu_{\mu\tau}$ fluxes, given by eq.(\ref{Fsk}).
For extracting the individual contributions of $\nu_e$ and $\nu_{\mu\tau}$,
i.e.:
\be
F_{a}(E_{\nu}) = 
\int dE'_{\nu} \; 
f_a(E'_{\nu}) \; \rho(\epsilon_e,E'_{\nu})    \;\;\;\;\;   (a=e,\,\mu,\,\tau)
\label{Fa}
\ee
additional information are needed.

\section{The information from $\nu_{e}+d\rightarrow p + p + e^{-}$}

SNO has recently presented \cite{sno} the energy spectrum of electrons 
from reaction  (\ref{cc_reac}). We show here that these data, which 
provide a direct determination of the electron neutrino spectrum, can
be combined with ES data for determining the spectrum of $\nu_{\mu}$
plus $\nu_{\tau}$.

 The SNO results are shown in the lower panel 
of fig.~\ref{fig4} as a function of the observed electron kinetic energy 
$T_{e} = \epsilon_{e} - m_{e}$. Data are grouped in 11 bins covering
the energy range $T_{e} = 6.75 - 13$ MeV. The signal in each bin, $C_{i}$,
is normalized to the predictions, $C_{i}^{SSM}$, corresponding to the SSM
of \cite{ssm} and the error bars take into account only statistical errors.

 The measured electron spectrum $C(\epsilon_{e})$ from reaction (\ref{cc_reac}),
normalized to the SSM expectation $C^{SSM}(\epsilon_{e})$, is related to the
the electron neutrino spectrum by:
\be
\frac{C}{C^{SSM}} = \int dE_{\nu} \; f_{e}(E_{\nu})\,
\rho_{cc}(\epsilon_{e},E_{\nu})
\label{cc_spec}
\ee
where $\rho_{cc}$ is a suitable response function, 
defined in terms of the cross section 
for reaction (\ref{cc_reac}), $d\sigma_{cc}/dE_{e}$ \cite{cross_sno},
 and of the detector resolution function 
$r_{\rm SNO}(\epsilon_{e},E_{e})$\footnote{
The SNO resolution function can be described by 
rel.~(\ref{resolution}) with 
$\Delta = (-0.4620+0.5470\,\sqrt{E_{e}}+ 0.008722\, E_{e})\;{\rm Mev}$ \cite{sno}. 
One can easily check that $r_{\rm SNO}\simeq r$.}:
\be
\rho_{cc}(\epsilon_{e},E_{\nu}) =
\frac{\displaystyle 
\varphi^{SSM}(E_{\nu}) \int dE_{e} \; 
r_{\rm SNO}(\epsilon_{e},E_{e}) \; 
\frac{d\sigma_{cc}(E_{\nu},E_{e})}{dE_{e}}
}
{\Psi_{cc}
}~.
\label{rho_cc}
\ee
The normalization factor $\Psi_{cc}$ is given by:
\be
\Psi_{cc} =
\int dE_{\nu} \;
\varphi^{SSM}(E_{\nu}) \int dE_{e} \;
r_{\rm SNO}(\epsilon_{e},E_{e}) \; 
\frac{d\sigma_{cc}(E_{\nu},E_{e})}{dE_{e}}~.
\label{psi_cc}
\ee

 The behaviour of the SNO response function is shown in fig.\ref{fig6_sno} 
as a function of neutrino energy, for representative $\epsilon_{e}$ values.
One sees that for each measured electron energy $\epsilon_{e}$ the
response function is peaked at a specific neutrino energy
$E_{\nu}$. The rates $C_{i}/C_{i}^{SSM}$ can thus be
 interpreted as a determination of the deviations of the
electron neutrino spectrum at $E_{\nu}$ averaged
over the response function $\rho_{cc}$, i.e.: 
\be
F_{\rm SNO}(E_{\nu})=
\int dE'_{\nu} \; f_{e}(E'_{\nu})
\rho_{cc}(\epsilon_{e},E'_{\nu})
\label{Fsno}
\ee
where the numerical relation $\epsilon_{e}(E_{\nu})$ is shown in
fig.~\ref{fig2}.

 We remark that the width of the SNO response function, 
$\rho_{cc}$, is mainly determined by the SNO resolution 
 $r_{\rm SNO}$.
In fact, if one neglects the recoil of the two protons in the
final state, there is a fixed relation between the neutrino and electron
energies, i.e. $d\sigma_{cc}/dE_{e} \propto \delta(E_{e}-E_{\nu}-m_{d}+2m_{p})$.
As a consequence one expects, roughly:
\be 
\rho_{cc}(\epsilon_{e},E_{\nu}) \propto 
\varphi^{SSM}(E_{\nu})\,r_{\rm SNO}(\epsilon_{e},E_{\nu}+m_{d}-2m_{p}).
\label{rho_cc_app}
\ee

We note that the structure of the previous relation is similar to that
of rel.~(\ref{rho}), which defines the response function $\rho$ associated to
the SK spectrum derivative. If one considers that SK and SNO energy
resolutions are almost equal,
this suggests that SK and SNO response 
functions can be equalized with a proper choice of 
the detection energies.

The qualitative argument given above is clearly oversimplified.
However, the possibility to equalize the SK and SNO response functions 
can be checked numerically. The results are shown in fig.~\ref{fig6}. One sees
that the response function $\rho_{e}$ and $\rho_{cc}$, 
calculated numerically according to rel.~(\ref{rho_a}) and (\ref{rho_cc}) 
are almost equal, i.e.
\be
\rho_{e}(\epsilon_{\rm SK},E_{\nu}) = \rho_{cc}(\epsilon_{\rm SNO},E_{\nu})
\label{sksno_equality}
\ee
if one chooses the energies as follows:
\be
\epsilon_{\rm SNO} = 0.975\,\epsilon_{\rm SK} - 2.50 \;{\rm MeV} ~.
\label{esno_esk}
\ee
 From the equalities~(\ref{rho_equality}) and (\ref{sksno_equality}),
 it follows that one can identify the SNO charged current spectrum
with the $\nu_{e}$ contribution to the SK spectrum derivative, i.e.:
\be
F_{e}(E_{\nu}) = F_{\rm SNO}(E_{\nu}) ~.
\label{Fe}
\ee
As a consequence,
 SK and SNO-CC data can be combined to determine:
\be
F_{\mu\tau}(E_{\nu}) = 
\frac{1}{\beta} \left( F_{\rm ES}(E_{\nu}) - F_{\rm SNO}(E_{\nu}) \right)
\label{Fmu}
\ee
and:
\be
F_{e}(E_{\nu}) + F_{\mu\tau}(E_{\nu}) = 
\frac{1}{\beta} \left[ F_{\rm ES}(E_{\nu}) - (1-\beta) F_{\rm SNO}(E_{\nu}) \right] ~.
\label{Ftot}
\ee
The results obtained are shown in fig.~\ref{fig7} where, for convenience,
the SNO-CC signals have been grouped into larger intervals so as to 
reduce statistical fluctuations and to take advantage of the SK-SNO
correspondence, eq.~(\ref{esno_esk}).

We remark that, if there are no sterile neutrinos, then the sum of the spectra
of active neutrinos has the same shape as the $^{8}$B spectrum in the 
laboratory, i.e.
\be
\varphi_{e}(E_{\nu})+\varphi_{\mu\tau}(E_{\nu}) =
k \; \varphi^{SSM}(E_{\nu})
\ee
where $k$ is a normalization constant, i.e. it does not depend on $E_{\nu}$.
Violations of this sum rule,
would clearly imply sterile neutrinos. Such deviations are not shown in the data,
which are well consistent with a constant, see the lowest panel of fig.\ref{fig7}.

\section{Concluding remarks}

We have provided a method for extracting information on the
neutrino spectrum from the spectrum of scattered electrons,
summarized in eq. (\ref{der_ratio}-\ref{ee_enu_app}). 
We summarize here a few points, concerning the possible
applications and developments of our approach:\\

{\it 1)} As an example we have applied our approach to the
published SK data, see fig.~\ref{fig5}. We suggest that the analysis
is performed by the experimental group, since a detailed 
knowledge of the detector is important for extracting optimal
information.\\

{\it 2)} We have shown that the information obtained by using our
method can be directly combined with the information provided
by SNO, so as to determine the spectrum of $\nu_{\mu}$ plus $\nu_{\tau}$
as well as the total spectrum of active neutrinos. 
Our results, obtained from eqs.~(\ref{Fmu}) and~(\ref{Ftot}),
 are shown in fig.~\ref{fig7}.\\

{\it 3)} We remark that the SK-SNO comparison potentially allows
for a model independent signature for sterile neutrinos.
If there are no sterile neutrinos, in fact, the sum of the spectra
of active neutrinos has the same shape as the $^{8}$B spectrum in the 
laboratory, i.e.
$
\varphi_{e}(E_{\nu})+\varphi_{\mu\tau}(E_{\nu}) =
k \; \varphi^{SSM}(E_{\nu})
$, 
where $k$ is a normalization constant. Violations of this sum rule,
would clearly imply sterile neutrinos. Such deviations are 
not shown in the data.\\

\section*{Acknowledgements}
G.F. is grateful to the CERN theory division for generous hospitality.

%..............................................................................

%..............................................................................
\newpage

\begin{figure}[htb]
\epsfig{file=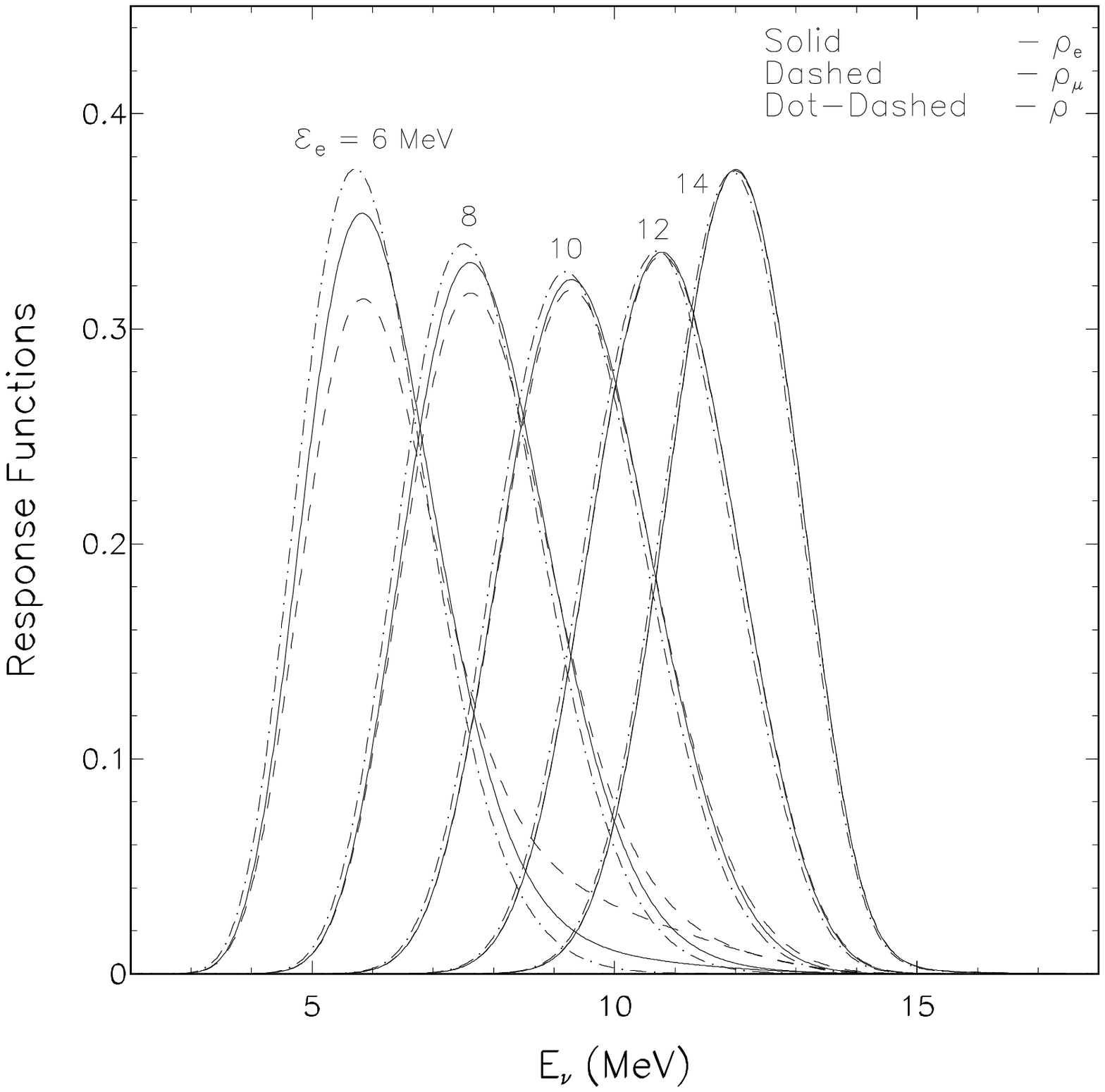,height=20cm,width=15cm}
\caption{Response functions of Super-Kamiokande 
as a function of neutrino energy $E_{\nu}$, for selected values of the 
observed electron energy $\epsilon_{e}$.
The exact response functions $\rho_{e}$ and $\rho_{\mu}$,
are calculated according to eq.(\ref{rho_a}), whereas $\rho$ is the
approximation given by eq.(\ref{rho}).}
\label{fig1}
\end{figure}

\begin{figure}[htb]
\epsfig{file=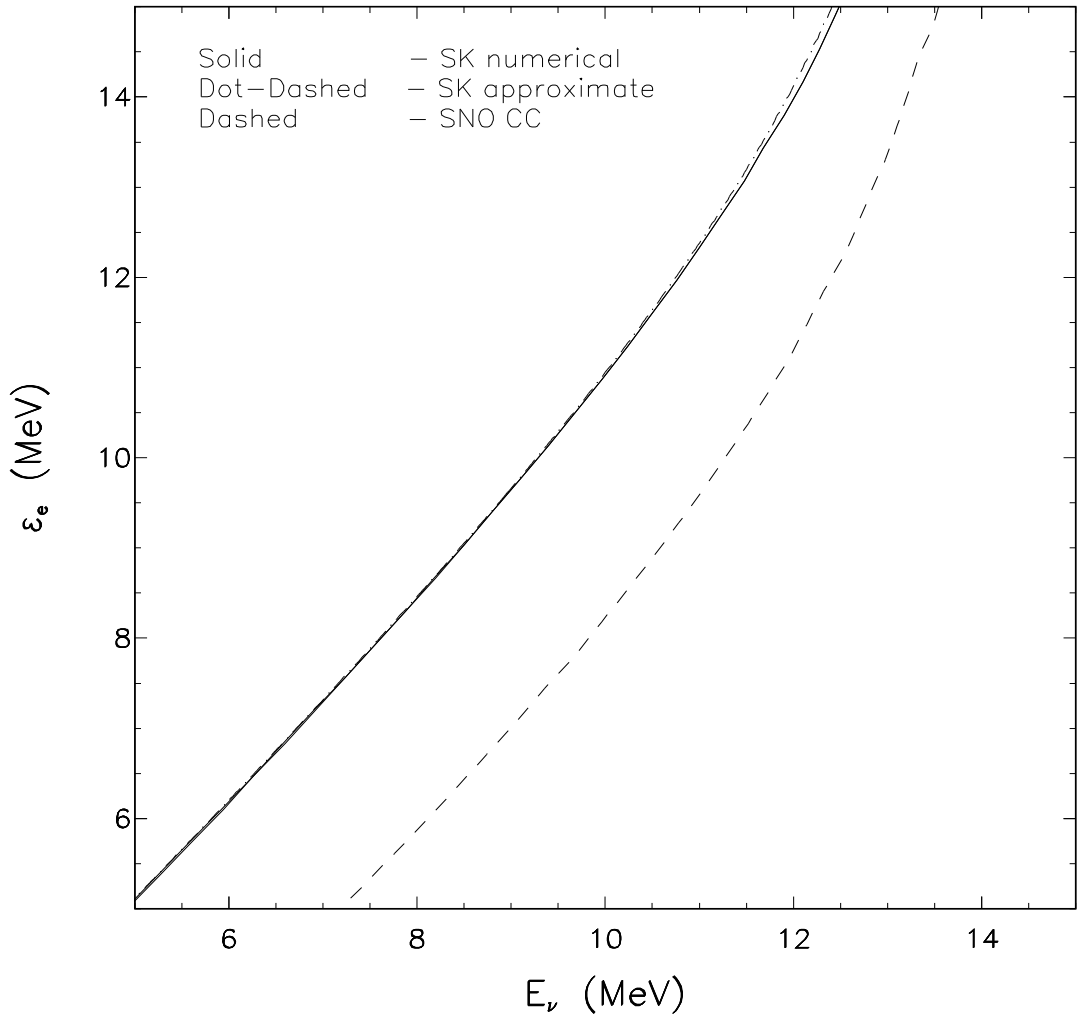,height=20cm,width=15cm}
\caption{The relation between neutrino and electron energies 
for SK and SNO detectors.
The upper lines refer to SK. They are obtained from numerical calculation (solid) 
and from the approximate relation (\ref{ee_enu_app}) (dot-dashed). The dashed line
refers to SNO.}
\label{fig2}
\end{figure}

\begin{figure}[htb]
\epsfig{file=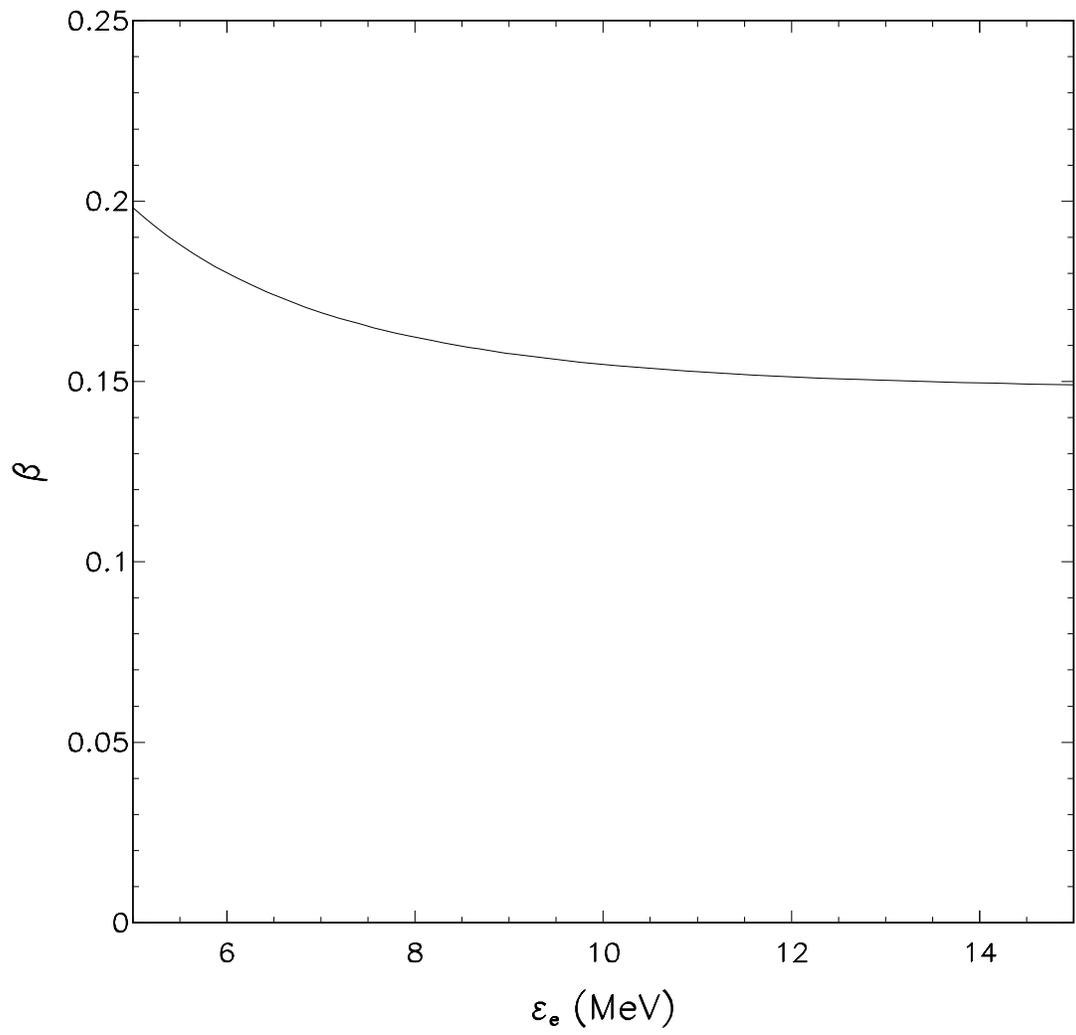,height=20cm,width=15cm}
\caption{The function $\beta$ from eq.(\ref{beta}).}
\label{fig3}
\end{figure}

\begin{figure}[htb]
\epsfig{file=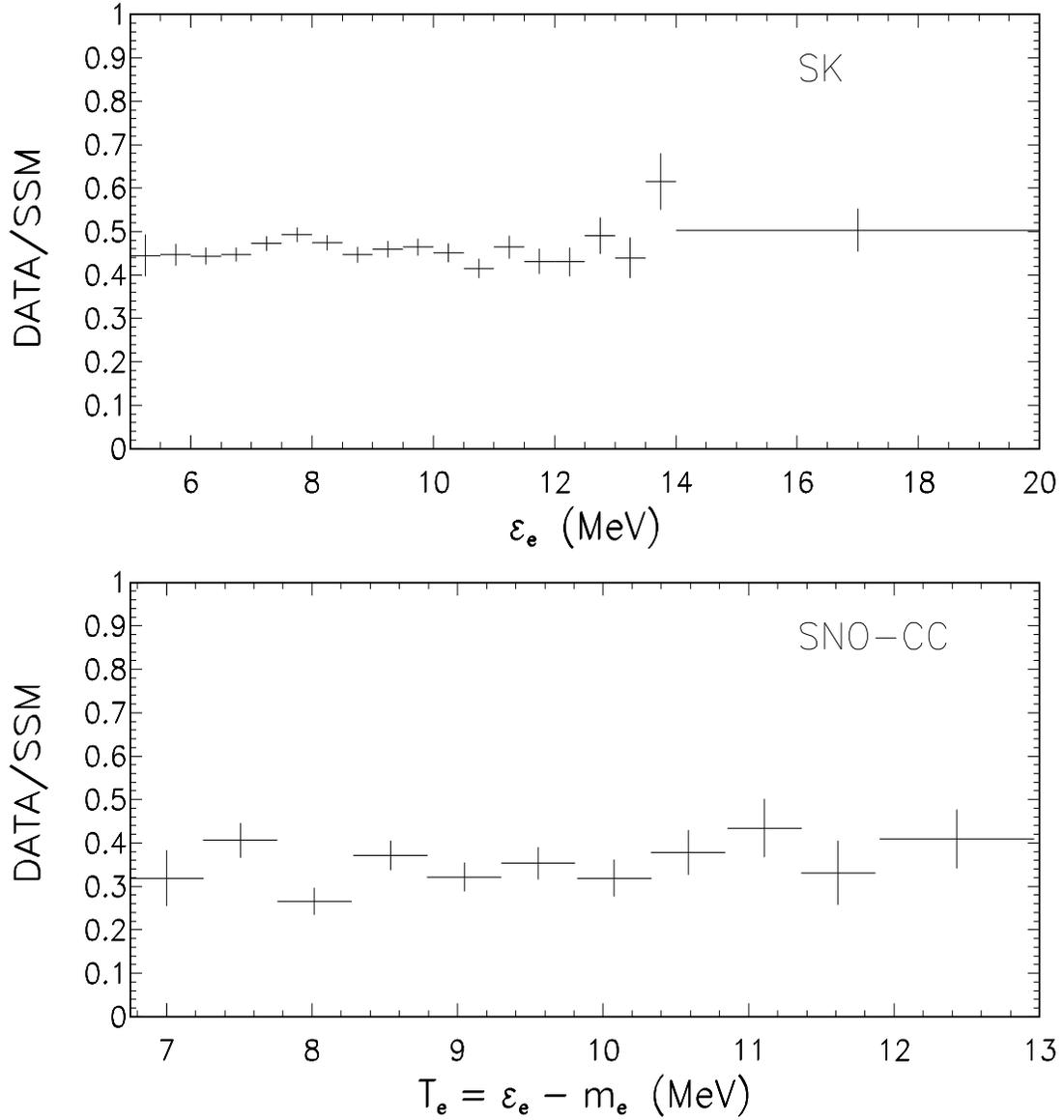,height=23cm,width=15cm}
\vspace{-3.4cm}
\caption{
Upper panel: The electron energy spectrum measured by SK \cite{sk}
normalized to the SSM prediction \cite{ssm}.
%, from \cite{sk}.\\
Lower panel: The electron energy spectrum measured by charged current
reaction (\ref{cc_reac}) at SNO \cite{sno}
normalized to the SSM prediction \cite{ssm}.}
%, from \cite{sno}.}
\label{fig4}
\end{figure}

\begin{figure}[htb]
\epsfig{file=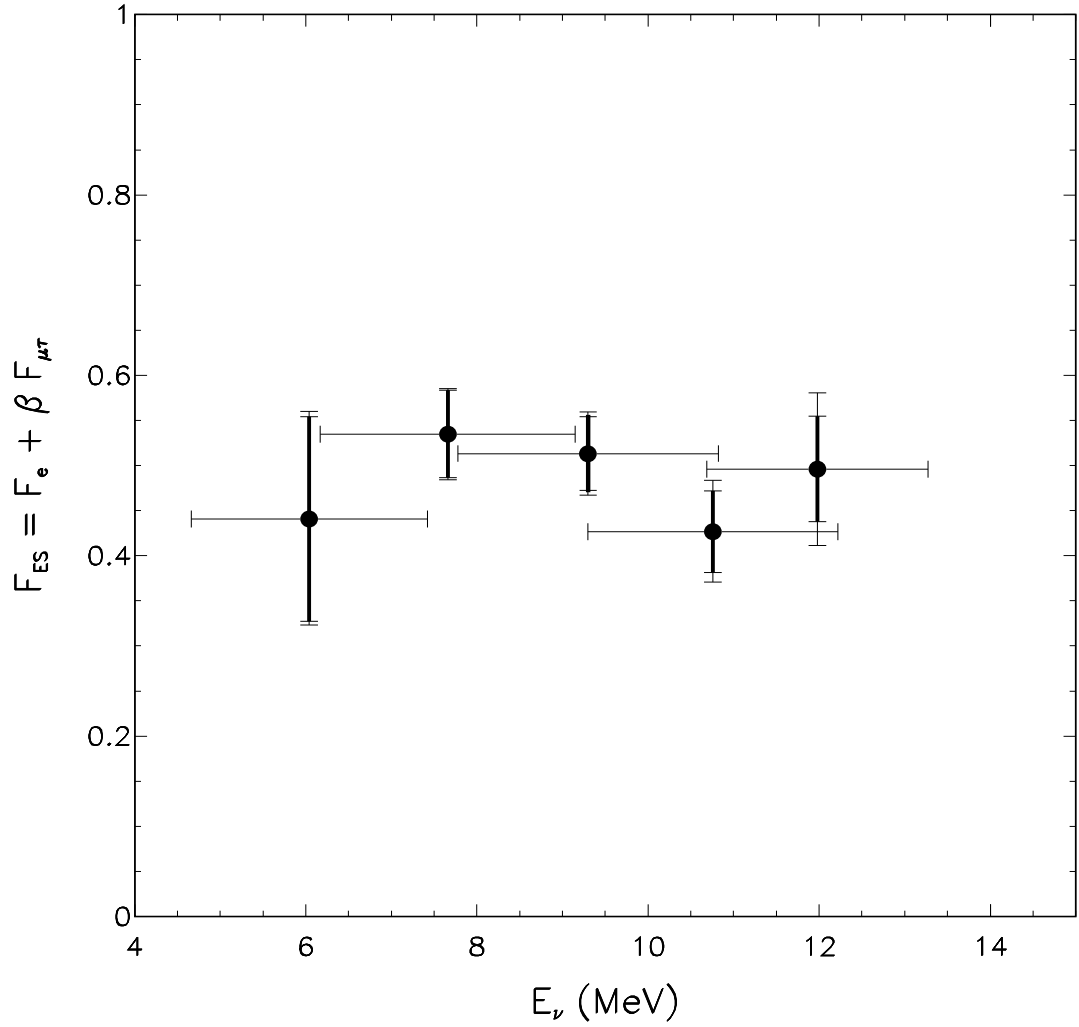,height=20cm,width=15cm}
\caption{Deviations of the neutrino spectrum from the SSM prediction, as a function of
neutrino energy $E_{\nu}$, from SK data.
The horizontal bar is the neutrino energy resolution. The inner vertical
bar denotes the uncorrelated error, the outer bar includes the effect of energy correlated
systematical errors.}
\label{fig5}
\end{figure}

\begin{figure}[htb]
\epsfig{file=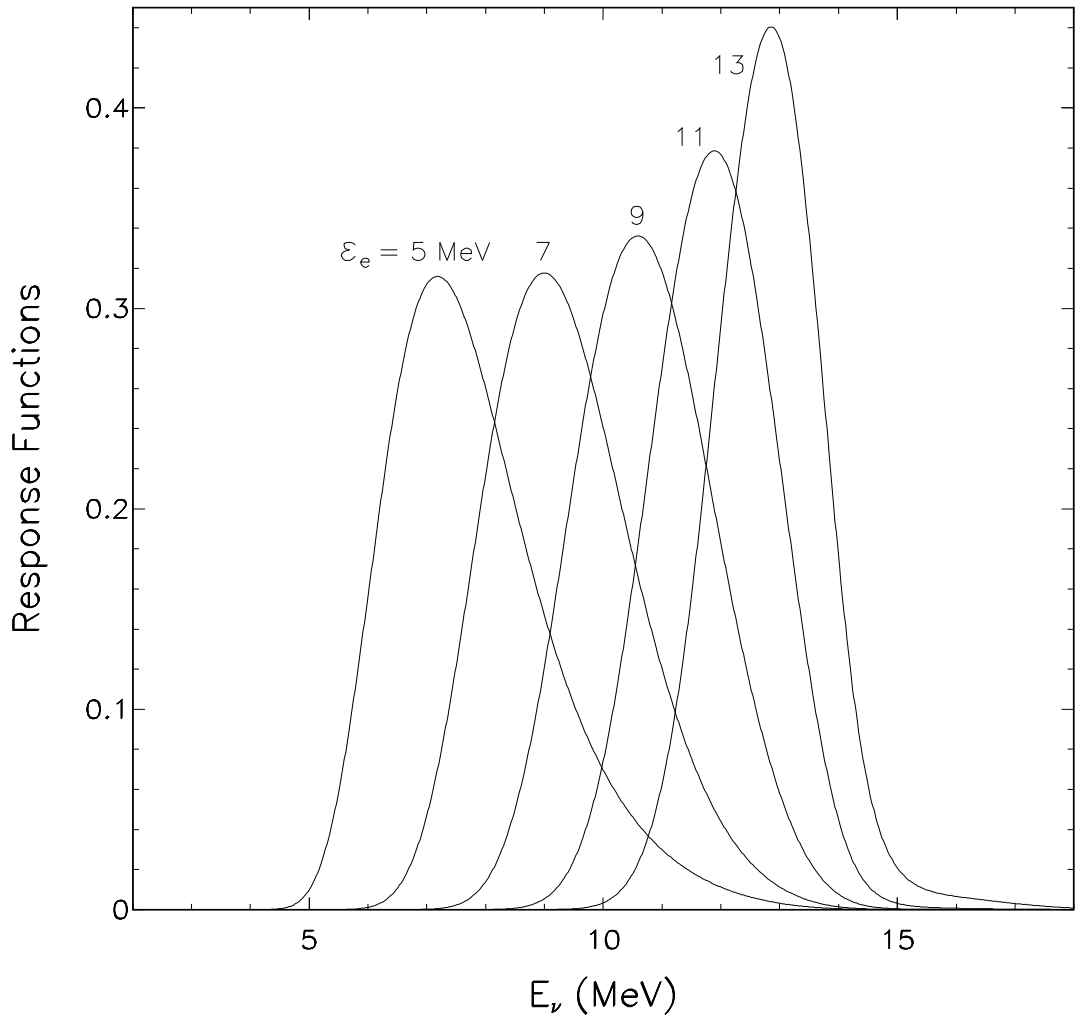,height=20cm,width=15cm}
\caption{The SNO charged current response function $\rho_{cc}$,
eq.(\ref{rho_cc}), as a function of neutrino 
energy $E_{\nu}$, for selected values of the 
observed electron energy $\epsilon_{e}$.}
\label{fig6_sno}
\end{figure}

\begin{figure}[htb]
\epsfig{file=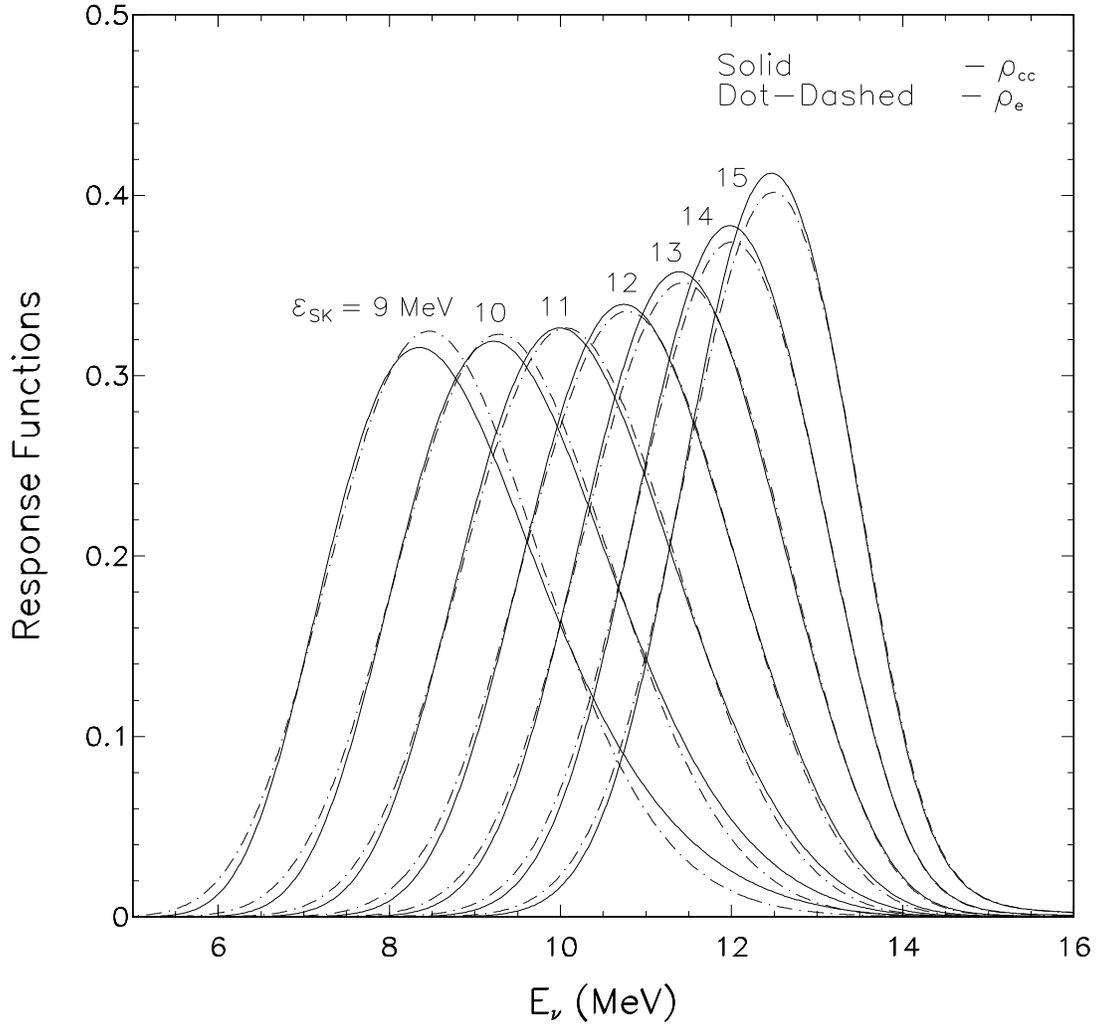,height=20cm,width=15cm}
\caption{
{\it a)} The SK response function $\rho_{e}$, eq.~(\ref{rho_a}), for the indicated electron
energies $\epsilon_{\rm SK}$ (dot-dashed lines).
{\it b)} The SNO response function $\rho_{cc}$, eq.~(\ref{rho_cc}), for the electron energies 
$\epsilon_{\rm SNO}$ given by eq.~(\ref{esno_esk}) (solid lines).}
\label{fig6}
\end{figure}

\begin{figure}[htb]
\epsfig{file=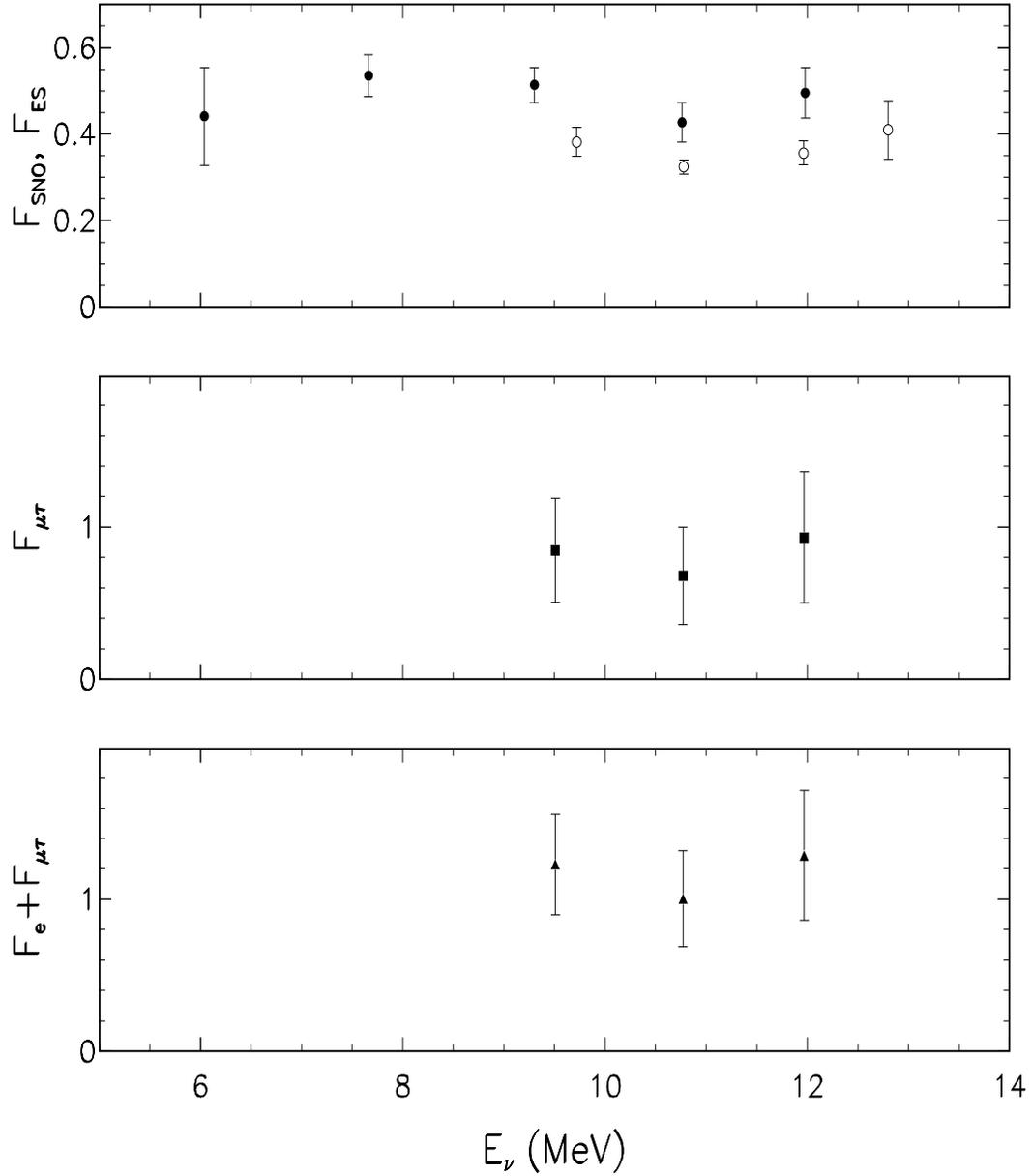,height=24cm,width=15cm}
\vspace{-3.5cm}
\caption{Deviations of the neutrino spectra from the SSM prediction as a 
function of the neutrino energy$ E_{\nu}$. The vertical bars take into account only
statistical errors. In the upper panel, the full circles correspond to 
$F_{\rm ES}$, while 
the open circles correspond to $F_{\rm SNO}$.}
\label{fig7}
\end{figure}

\end{document}